\documentclass[conference]{IEEEtran}
\IEEEoverridecommandlockouts
\usepackage{cite}
\usepackage{amsmath,amssymb,amsfonts}
\usepackage{algorithmic}
\usepackage{graphicx}
\usepackage{textcomp}
\usepackage{xcolor}
\usepackage{todonotes}
\usepackage{cleveref}
\usepackage{caption}
\usepackage{pifont}
\usepackage{booktabs}
\usepackage{url}
\def\BibTeX{{\rm B\kern-.05em{\sc i\kern-.025em b}\kern-.08em
    T\kern-.1667em\lower.7ex\hbox{E}\kern-.125emX}}
    
\begin{document}

\title{In Serverless, OS Scheduler Choice Costs Money: A Hybrid Scheduling Approach for Cheaper FaaS\thanks{Author draft made available for timely dissemination.}}

\author{\IEEEauthorblockN{Yuxuan Zhao}
\IEEEauthorblockA{LIACS, Leiden University \\
y.zhao@liacs.leidenuniv.nl}
\and
\IEEEauthorblockN{Weikang Weng}
\IEEEauthorblockA{LIACS, Leiden University \\
w.weng@liacs.leidenuniv.nl}
\and
\IEEEauthorblockN{Rob van Nieuwpoort}
\IEEEauthorblockA{LIACS, Leiden University \\
r.v.van.nieuwpoort@liacs.leidenuniv.nl}
\and
\IEEEauthorblockN{Alexandru Uta}
\IEEEauthorblockA{DFINITY, Zurich \\
alexandru.uta@gmail.com}
}

\maketitle

\begin{abstract}
In Function-as-a-Service (FaaS) serverless, large applications are split into short-lived stateless functions. Deploying functions is mutually profitable: users need not be concerned with resource management, while providers can keep their servers at high utilization rates running thousands of functions concurrently on a single machine. It is exactly this high concurrency that comes at a cost. The standard Linux Completely Fair Scheduler (CFS) switches often between tasks, which leads to prolonged execution times. We present evidence  that relying on the default Linux CFS scheduler increases serverless workloads cost by up to $10\times$.

In this article, we raise awareness and make a case for rethinking the OS-level scheduling in Linux for serverless workloads composed of many short-lived processes. To make serverless more affordable we introduce a hybrid two-level scheduling approach that relies on FaaS characteristics. Short-running functions are executed in FIFO fashion without preemption, while longer-running functions are passed to CFS after a certain time period. We show that tailor-made OS scheduling is able to significantly reduce user-facing costs without adding any provider-facing overhead.
\end{abstract}


\section{Introduction}
For Function-as-a-Service (FaaS) serverless, applications are sliced into short-lived functions, allowing cloud providers to dynamically allocate resources and manage server provisioning in a fine-grained manner on behalf of the user~\cite{jonas_cloud_nodate, roberts2017serverless}. 
The fine-grained and elastic scaling characteristics allow cloud providers to have profit margins by ultimately taking advantage of packing many functions into the same server for high utilization. From the user perspective, serverless applications are scalable and cost-effective~\cite{rohit2018building, copik2021sebs}. Users do not need to configure, manage, and maintain servers, they just ship code or container images~\cite{brooker2023demand} to the platform and receive results back~\cite{van_eyk_spec_2017,jonas_occupy_2017}.

Serverless functions exhibit runtime skew and invocation burstiness~\cite{ginzburg2020serverless, kaffes2019centralized, fuerst2022locality, talluri2023trace, eismann2020review}. According to a Microsoft Azure study~\cite{shahrad2020serverless}, 80\% of the serverless functions execute less than 1 second (see the cumulative distribution function plot in Figure~\ref{spike}), more than 90\% of functions allocate virtual memory less than 400MB, and 81\% of functions are invoked once per minute or less. Since the execution of a wide majority of serverless functions is typically short, on the order of seconds or less~\cite{shahrad2020serverless,fuerst2022locality, zhao2022tiny, zhang2021faster, yu2020characterizing}, cloud providers charge users according to their function execution duration with a price for every 1 millisecond. Costs are also related to the amount of memory allocated to the user's functions~\cite{aws_pricing, eismann2021sizeless, elgamal2018costless, eivy2017wary}. 

When handling serverless workloads, cloud providers aim at low latency and execution time for the users, while aiming at high resource utilization on the provider side to achieve economies of scale~\cite{li2022rund}. For instance, AWS Lambda is thought to be able to deploy 4$\sim$8 thousands of instances on a single host machine~\cite{ustiugov2021benchmarking,agache2020firecracker}. This shows that FaaS would not be cost-effective without massive CPU sharing between users (i.e., concurrent functions), which means noisy neighbors and scheduling influence overall runtime and cost. The cloud provider charges the user for wall clock time instead of CPU time. In AWS Lambda~\cite{AmazonLambda}, for example, billing for wall clock time, users are also billed for time spent waiting for external resources such as databases. For example, if a function is actively running on CPU for 1 millisecond and waiting 1 minute for an external database to return a query, AWS Lambda will bill for the whole 1 minute, not just the 1 millisecond CPU time.

Recent studies~\cite{fu2022sfs,tian2022owl,li2023golgi,kaffes2022hermod,kaffes2019centralized,zuk2022call,isstaif2023towards} indicate that scheduling for serverless functions impacts the efficiency of FaaS systems. Given the characteristic of serverless functions, where the majority of the invocations are very short, on the order of seconds or less, the \textit{de facto} standard Linux scheduler, the Completely Fair Scheduler (CFS) may lead to performance degradation due to context switches. Such frequent context switches introduce unnecessary overhead on execution times, as a consequence of costly state saving and restoration~\cite{humphries2021case}. 

\begin{figure}[tp]
    \centering
    \includegraphics[width=0.95\linewidth]{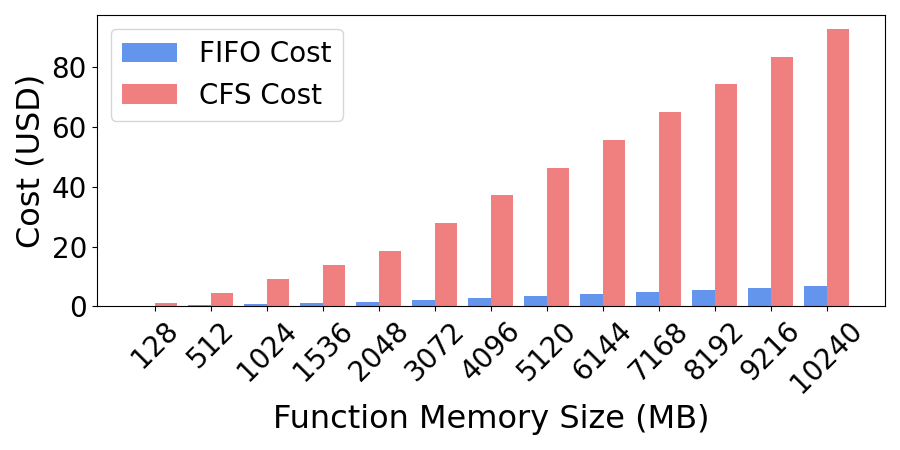}
    \caption{Cost for FIFO and CFS OS scheduling policies calculated using AWS Lambda pricing. The workload is using the first 12,442 functions in the Microsoft Azure trace. Although FIFO cost is significantly lower, it introduces unacceptably large latencies for functions that simply wait in queues. We explore this trade-off in the remainder of the article.}
    \label{cost gap fifo}
    \vspace*{-0.5cm}
\end{figure}

CFS aims to ensure fairness by assigning a minimum CPU time slice to every running task~\cite{lozi2016linux}. However, CFS might not be optimal for serverless function scheduling~\cite{fu2022sfs,isstaif2023towards,koller2017will}. This is because it does not address the special nature of serverless computing, where many thousands of short-running tasks are running concurrently. A previous study~\cite{isstaif2023towards} modifies CFS to give a higher priority to long-tail functions and achieves a $5{\sim}30\%$ reduction in latency. However, this is insufficient for significant cost reduction because CFS-induced longer running times lead to larger user-facing cost. Figure~\ref{cost gap fifo} provides evidence for how much extra cost is incurred via CFS in comparison to FIFO for different function memory sizes. For these first 12,442 function invocations of the Azure trace~\cite{shahrad2020serverless}, CFS introduces more than 10 times extra cost compared to FIFO. It is nevertheless true that FIFO is also suboptimal because it increases queueing time and hence user-facing latency significantly. We therefore raise awareness and make a case for rethinking OS-level scheduling for serverless functions.

To explore the latency-cost trade-off, we implement a hybrid approach for serverless function scheduling by separating the scheduler into two distinct groups of CPU cores using ghOSt~\cite{humphries2021ghost}, a scheduling policy delegation system in user space. A group of CPUs are specialized for rapidly processing short-lived serverless functions. Long-running functions will be migrated from these specialized CPUs after a certain time limit to leave room for other short tasks. If functions finish before running out of time, they have been running without any interruption by context switches, reducing their execution time and cost. Functions not completed in the time limit are preempted and scheduled on another group of CPUs, designated for handling long-running functions (e.g., longer than 1 second). We evaluate our hybrid scheduler in two different modes of running functions: regular Linux processes and AWS Lambda invocations~\cite{AmazonLambda,agache2020firecracker}. This covers both common serverless operating use-cases: using containers and using (micro-) virtual machines. In our experiments, we showcase how this approach decreases the number of context switches for the short-lived functions and leads to a shorter duration, hence less user-facing cost. This comes at no additional overhead for the provider. In this paper, we raise awareness for the fact that scheduling in serverless is by far not a solved problem. We thereby present an initial solution for reducing cost in serverless via OS-level scheduling.

Furthermore, our hybrid scheduler introduces a mechanism for adapting the FIFO time limit based on recent past function executions. Additionally, to avoid the CPUs being under-utilized, we introduce a mechanism that dynamically adapts the number of CPU cores designated for the two scheduling policies. We monitor the CPU cores utilization and move cores from one group to another if needed. In this way, better load balance and CPU utilization is achieved for both groups of CPU cores, 
which adheres to the interests of cloud infrastructure providers.

In summary, we make the following contributions:
\begin{enumerate}
    \item We showcase evidence that CFS is sub-optimal for serverless functions and increases the cost for serverless customers (Section~\ref{motivation}).
    \item We design and implement a hybrid scheduler in user space using Google ghOSt~\cite{humphries2021ghost} that divides the CPU cores into two groups and both groups are designated a separate scheduling policy. The function will be preempted from one core group to the other if it can not finish in a specific time limit (Section~\ref{design}).
    \item We design and implement a time limit adaptation and a rightsizing mechanism for the number of CPU cores to keep CPU utilization high (Section~\ref{design}).
    \item We showcase empirically that our scheduler improves the execution time of serverless functions as well as reduces cost on real-world platforms, such as Firecracker ~\cite{agache2020firecracker} (Section~\ref{evaluation}).
\end{enumerate}

\section{Motivation}
\label{motivation}
In this section, we illustrate why the scheduling policy is important for serverless functions and why the Linux default CFS policy is suboptimal in this case. In a previous study~\cite{isstaif2023towards}, the authors modify CFS and give an in-depth discussion on kernel-space scheduling and its limitations. Therefore in this work, we explore user-space scheduling. We first introduce the characteristics of serverless functions and the usual price model from popular cloud platforms~\cite{AmazonLambda, AzureFunctions, GoogleFunction} (Section~\ref{FaaS Characteristics}). Then we introduce the metrics used in this paper (Section~\ref{Metrics}). We compare the two most well-known scheduling policies, CFS and FIFO (and its variant), on 50 Intel Xeon CPU cores (Section~\ref{config}) while running the first 12,442 functions in the Azure trace~\cite{shahrad2020serverless} (Section~\ref{CFS vs. FIFO}).

Furthermore, we also showcase why we need preemption for the serverless functions (Section~\ref{Why do we need preemption?}). Then we showcase why we need to separate CPU cores into different scheduling policies (Section~\ref{Why do we need hybrid OS scheduling?}). Our observations demonstrate the FIFO policy can achieve near-optimal execution time of serverless functions with poor response time (i.e., high user-facing latency). In terms of the default Linux CFS policy, it reduces the response time of serverless functions significantly, but however it introduces prolonged execution times. As a consequence, this results in unnecessary extra costs for serverless users (Section~\ref{How large is the CFS cost increase?}).

\textbf {Workload.} The workload used throughout this section is represented by the first 12,442 functions in the Microsoft Azure FaaS trace (i.e., the first two minutes). We extract the inter-arrival pattern from the Azure FaaS trace and downscale the number of functions. The time when the function is created depends on the inter-arrival time. For more details about the workload see Section~\ref{workload generator}.

\subsection{FaaS Characteristics}
\label{FaaS Characteristics}
\begin{figure}[htp]
    \centering
    \begin{minipage}[b]{0.42\linewidth}
        \centering
        \includegraphics[width=\linewidth]{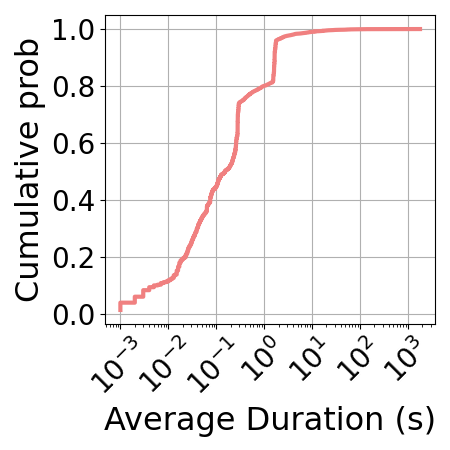}
    \end{minipage}
    \hfill
    \begin{minipage}[b]{0.55\linewidth}
        \centering
        \includegraphics[width=\linewidth]{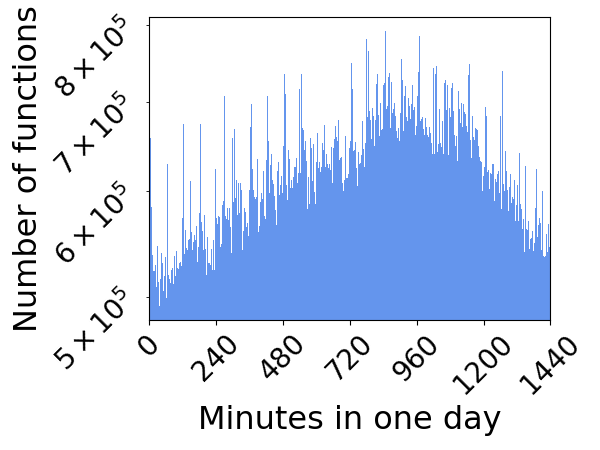}
    \end{minipage}
        \caption{\textbf{Left:} Average duration distribution in a two-week period in Azure dataset. \textbf{Right:} Function arrival pattern on the first day in Azure dataset, noting the burstiness characteristic.}
    \label{spike}
\end{figure}

Serverless functions are typically characterized by their burstiness and short-lived features~\cite{fuerst2022locality}. As depicted in Figure~\ref{spike}, the serverless arrival pattern exhibits sudden spikes. This phenomenon indicates that cloud providers need to make quick adaptations to allocate resources efficiently to handle such burst arrival patterns. 

Furthermore, according to the Azure study~\cite{shahrad2020serverless}, serverless functions are typically short-lived, the majority (80\%) complete their executions in less than 1 second as Figure~\ref{spike} illustrates. Because of the short-lived nature of the serverless function, scheduling overheads significantly influence performance and costs.

As mentioned before, cloud provider like AWS Lambda~\cite{AmazonLambda} charges users a per-millisecond price. The per-millisecond price also depends on the memory size allocated to the serverless functions~\cite{aws_pricing, eismann2021sizeless, elgamal2018costless, eivy2017wary}. A higher allocated memory size has a higher per-millisecond price but results in a faster execution time. There is a trade-off between the allocated memory size and incurred cost. For an in-depth discussion, we refer to~\cite{eismann2021sizeless}. To focus on the effects of scheduling policies, we compare the serverless function costs with a fixed memory sizes. Moreover, due to security concerns, just like serverless providers, we do not treat the case when multiple functions share the same sandbox. This would open up possible attacks that break proper separation of resources between functions, leading to data leaks.\\

\noindent\fbox{%
    \parbox{\linewidth}{%
        \textbf{Observation 1:} Because of their short-lived nature and their "pay per millisecond" pricing model, serverless functions are very interruption-sensitive.
    }%
}

\subsection{Metrics}
\label{Metrics}
\begin{figure}[htp]
    \centering
    \includegraphics[width=0.9\linewidth]{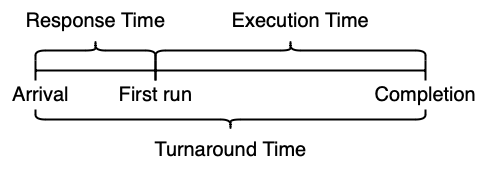}
    \caption{Metrics from~\cite{ArpaciDusseau23-Book}. Note that the more tasks accumulate in the global queue, the response time for each task tends to increase. Additionally, since the task is preempted, the execution time will be extended until the task is complete. }
    \label{fig:metrics}
    \vspace*{-0.4cm}
\end{figure}

To compare scheduling policies, we use the metrics presented in~\cite{ArpaciDusseau23-Book}, which include execution time, response time, and turnaround time as follows (see Figure~\ref{fig:metrics}). The execution time indicates the duration from the task's first run on the CPU to the time when the task is finished. As tasks are preempted during execution, they are suspended temporarily before being resumed. This preemption extends the overall execution time, directly impacting the cost for serverless users. The response time refers to the duration from the task's arrival to the time when the task first runs. When many tasks accumulate in the global queue, the response time for each task tends to increase due to head-of-line blocking. Head-of-line blocking negatively affects the response time. The turnaround time is the whole time the task was in the system, including execution time and response time.

As shown in Figure~\ref{fig:metrics}, these metrics can offer insights into the behavior of tasks under different scheduling policies in a serverless environment, particularly regarding how preemptions and task queueing can influence overall performance.

$$
T_{\text {execution }}=T_{\text {completion }}-T_{\text {firstrun }},
$$
$$
T_{\text {response }}=T_{\text {firstrun }}-T_{\text {arrival }},
$$
$$
T_{\text {turnaround }}=T_{\text {completion }}-T_{\text {arrival }}.
$$

\subsection{CFS vs. FIFO}
\label{CFS vs. FIFO}
\begin{figure}[htp]
    \centering
    \includegraphics[width=0.99\linewidth]{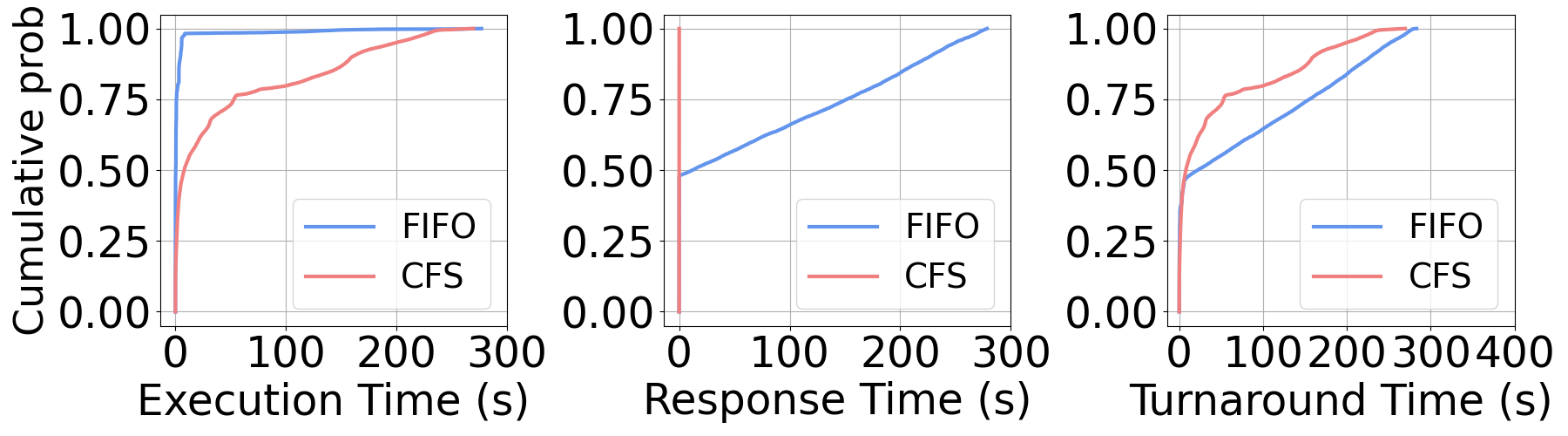}
    \caption{Metrics comparison between FIFO and CFS. FIFO policy achieves good execution time but sacrifices latency.}
    \label{FIFOvsCFS}
\end{figure}

We first compare two of the most well-known scheduling policies, FIFO and CFS. Figure~\ref{FIFOvsCFS} illustrates this comparison. The execution time comparison demonstrates most of the serverless functions have less execution time with FIFO than with CFS because of significantly fewer task preemptions and context switches. 

However, since CFS pursues fairness, the frequent context switch ensures the waiting time of tasks in the queue is little, which leads to a nearly vertical CDF line in Figure~\ref{FIFOvsCFS}. With FIFO, head-of-line blocking unavoidably hampers response time, resulting in a worse overall turnaround time.\\

\noindent\fbox{%
    \parbox{\linewidth}{%
        \textbf{Observation 2:} For short-lived functions, OS schedulers significantly influence performance. For example, FIFO is significantly better than CFS at  execution time, but also significantly worse at response time, leading to possible trade-offs between execution time (or cost) vs. user-facing latency.
    }%
}

\subsection{Why do we need preemption?}
\label{Why do we need preemption?}
\begin{figure}[htp]
    \centering
    \includegraphics[width=0.99\linewidth]{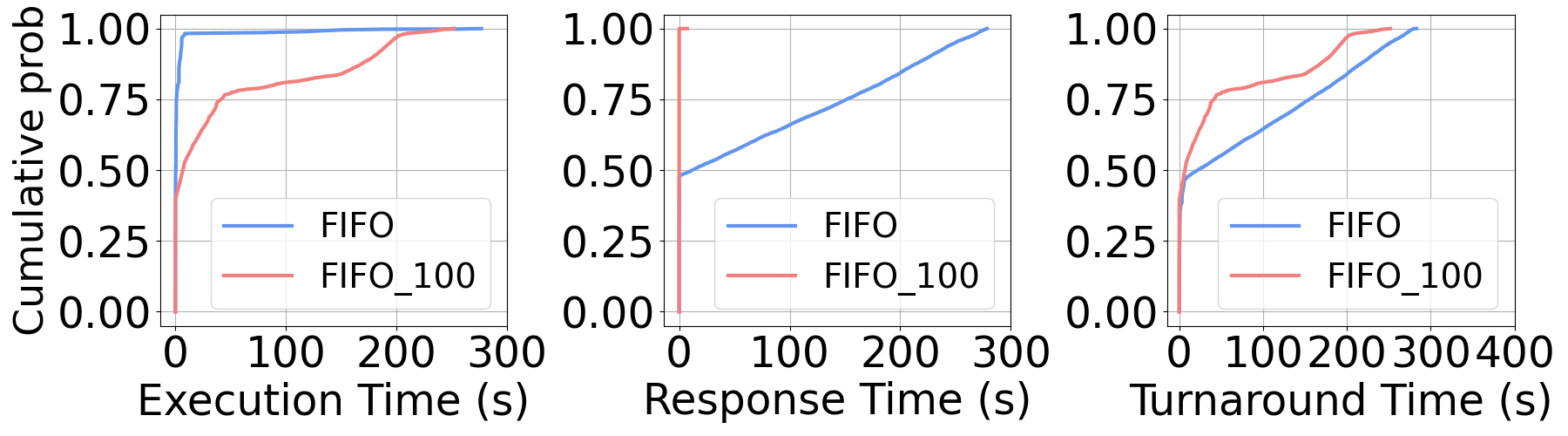}
    \caption{Metrics comparison between FIFO policy and FIFO policy with 100ms preemption. Preemption improves response time at the cost of increasing execution time.}
    \label{FIFOvsFIFO_100ms}
\end{figure}

Although the FIFO policy can achieve good execution time, the response time of the tasks suffers because of head-of-line blocking. If the task is preempted and moved to the end of the global queue, the head-of-line blocking problem in the global queue is somewhat alleviated. We implemented a custom version of FIFO, denoted FIFO\_100ms in which the task will be preempted and moved to the end of the queue if it exceeds running 100 ms.
Figure~\ref{FIFOvsFIFO_100ms} shows the comparison between FIFO and FIFO\_100ms. For the latter, response times improve significantly although the execution time increases. However as a consequence, the overall turnaround time still improves compared to plain FIFO.\\

\noindent\fbox{%
    \parbox{\linewidth}{%
        \textbf{Observation 3:} Preempting tasks and moving them to the end of the queue can significantly improve response time at the cost of increased execution time. Although the execution time deteriorates, the overall turnaround can still improve.
    }%
}

\subsection{Why do we need hybrid OS scheduling?}
\label{Why do we need hybrid OS scheduling?}
\begin{figure}[htp]
    \centering
    \includegraphics[width=0.99\linewidth]{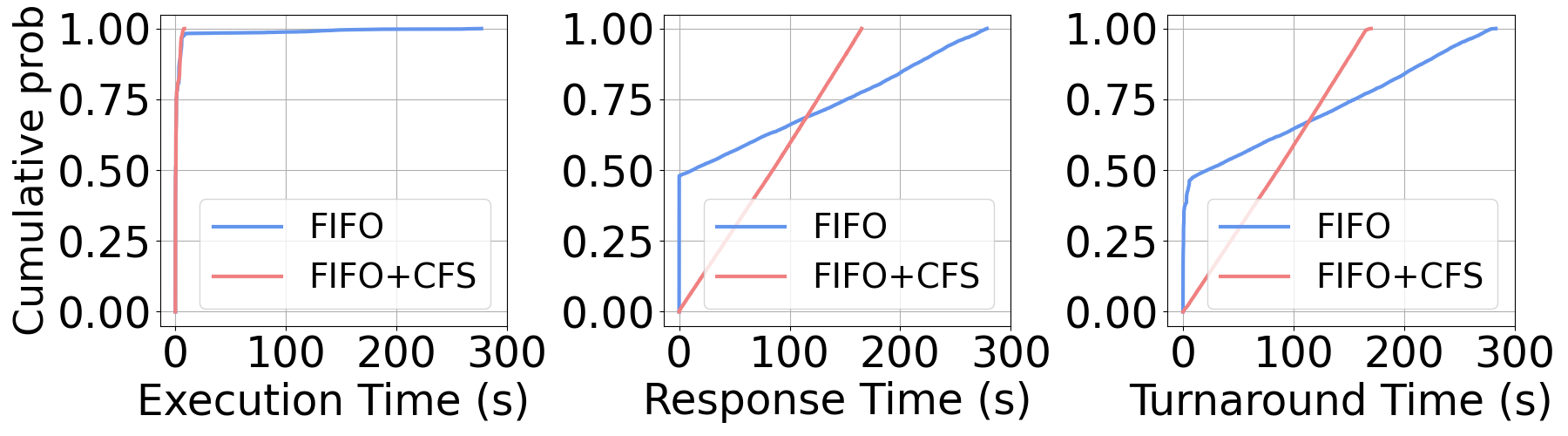}
    \caption{Metrics comparison between FIFO policy and hybrid FIFO and CFS policies designated to two CPU core groups. Our hybrid policy improves execution time, response time, and also turnaround time.}
    \label{FIFOvsFIFO+CFS}
\end{figure}

Since there is a trade-off between execution time and response time, an intuitive idea is to designate several cores to process only the tasks that are preempted from the FIFO policy. With this approach, we have the opportunity to improve response time while reducing the impact on execution time. Figure~\ref{FIFOvsFIFO+CFS} shows the comparison between FIFO and FIFO+CFS. FIFO+CFS refers to the separation that divides the CPU cores into two groups, one group is designated with the FIFO policy, and the other is designated with the CFS policy. Tasks initially enter the FIFO policy. If a task is not completed in a specific time period, it is preempted to cores running the CFS policy. Specifically, we separate 25 cores deploying the FIFO policy and 25 cores deploying the CFS policy. Additionally, we set the FIFO time limit to 1,633 ms (the 90th percentile of function duration in our workload). The end result is that the hybrid OS scheduling improves all the metrics of the FIFO policy. In the remainder of the article we show how to refine these results further.

\vspace*{0.4cm}
\noindent\fbox{%
    \parbox{\linewidth}{%
        \textbf{Observation 4:} A hybrid scheduler that treats short and long tasks with separate OS scheduling policies and on separate sets of cores improves all metrics.
    }%
}

\subsection{How large is the CFS cost increase?}
\label{How large is the CFS cost increase?}
Since under CFS tasks are kept time-slicing to ensure fairness, the execution time of each task is unavoidably prolonged. As a consequence of the extended execution time, serverless users face higher costs. As Figure~\ref{cost gap fifo} depicts, functions under the CFS policy incur more than 10 times extra cost than under FIFO. The cost is calculated by multiplying the total execution time of these 12,442 functions by the cost per millisecond from AWS lambda pricing~\cite{aws_pricing}.\\


\noindent\fbox{%
    \parbox{\linewidth}{%
        \textbf{Observation 5:} For short-running serverless functions and their per-millisecond pricing, prolonged execution time due to CFS leads to up to 10X extra user-facing costs.
    }%
}

\section{Background}
As Linux kernel development is considered difficult, error prone, and has long iteration times, many schedulers turn to Linux user space development~\cite{ford1996cpu, kaffes2021syrup, qin2018arachne, humphries2021ghost, fu2022sfs}. In this paper, we decide to implement our scheduler on ghOSt~\cite{humphries2021ghost}, which is a state-of-the-art scheduler delegation system in Linux user space. We introduce the design of the ghOSt system and several schedulers implemented in ghOSt.

\subsection{ghOSt design}
ghOSt~\cite{humphries2021ghost} is a scheduling policy delegation system developed by Google for handling kernel scheduling decisions. As the demand for data center workloads and platforms continues to grow at a rapid pace, there is an increasing need to customize scheduling policies to meet specific performance metrics like latency, load balancing, and energy efficiency. However, customizing and fine-tuning scheduling policies in the Linux kernel can be a complex and intricate task. 
ghOSt enables users to develop and optimize their own scheduling policies, tailored to their specific requirements, all without the need for a host machine reboot. Because ghOSt is able to delegate the scheduling policy to the user side and schedule the native OS threads. Even though ghOSt works as a delegation system, it manages to deliver throughput and latency that are on par with the performance of the kernel scheduler.

The ghOSt infrastructure consists of two components: a kernel module and its user space delegation mechanism. ghOSt scheduling class is on the kernel side and is responsible for providing APIs. These APIs are used to expose the state of the scheduling threads. The changes in the thread states are passed to the ghOSt agents on the user space side. ghOSt agents run in user space and are responsible for making scheduling decisions. The ghOSt agents pass the scheduling decisions to the kernel side through system calls. 

ghOSt introduces enclaves, which consist of multiple CPU cores. Each enclave can run its own scheduling policy. Through eclaves, ghOSt is capable of partitioning the machine at the CPU core granularity. Developers are responsible for implementing (in user space) the scheduling policies for different enclaves. 

\subsection{Why not Linux schedtool?}
schedtool~\cite{schedtool} is a Linux utility for handling 
CPU scheduling policies of the running processes. In contrast, ghOSt adapts scheduling by introducing a concept called enclave which is a group of CPU cores and deploys a custom scheduling policy to that enclave. schedtool adapts the scheduling policy by changing the scheduling parameters of each process. More importantly, in terms of schedulers, ghOSt enables customized schedulers but Linux schedtool only supports scheduling policies available in the kernel, such as FIFO, RR, BATCH, and the default CFS.

\subsection{Scheduling policies}
First in, First Out (FIFO) is the most basic scheduling policy. FIFO schedules tasks in the order they arrive, and the tasks run to completion without any preemption. Because there are no interruptions when the task is running, FIFO achieves the optimal execution time. However, FIFO is sensitive to head-of-line-blocking, increasing latency.

Completely Fair Scheduler (CFS) is a preemptive scheduling policy, the Linux default. Tasks are given CPU time on a time-slicing basis. Each CPU core maintains a run queue. CFS assigns each task a virtual runtime (vruntime), which is monotonically increasing. The task with the smallest vruntime is selected to execute next. To give higher priority tasks more CPU time, CFS adjusts the vruntime more slowly for the higher priority tasks.

Earliest Deadline First (EDF) is a scheduling policy mainly used in real-time systems, where tasks must finish in a certain time limit. EDF prioritizes tasks based on their deadlines and the task with the nearest deadline is always given high priority and executed first. EDF policy is a type of preemptive policy where running tasks can be interrupted if a new task with a closer deadline arrives.

Round Robin is a preemptive scheduler that gives CPU tasks a time slice for execution. Round Robin selects the first task in the queue to execute next. If the task runs out of the time slice but has not been completed yet, Round Robin interrupts the task and resumes the task the next time a time slice is assigned to that task.

Shinjuku~\cite{kaffes2019shinjuku} is a preemptive scheduler that can realize fast preemption at milliseconds scale. Shinjuku achieves truly centralized scheduling through one or more dedicated threads with a centralized view of load distribution. Furthermore, Shinjuku takes advantage of features of x86 hardware virtualization to minimize preemption overheads. Shinjuku significantly improves the tail latency and throughput.

\begin{figure}[tp]
    \centering
    \includegraphics[width=0.99\linewidth]{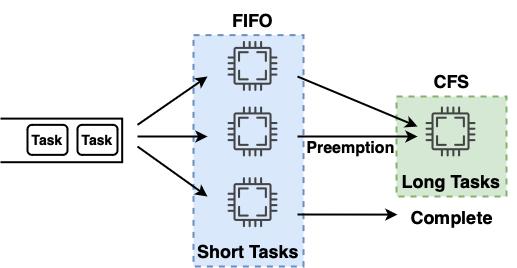}
    \caption{Our scheduler architecture. Task are first directed to the FIFO policy. If the task completes execution before the time limit, it finishes without any interruptions. Otherwise, it will be preempted to the cores which deploy the CFS policy to continue the remain execution.}
    \label{scheduler architecture}
\end{figure}

\section{Hybrid Scheduler Design}
\label{design}
We introduce our hybrid scheduler design, the mechanism to dynamically adapt the FIFO time limit, and the mechanism for the CPU core group rightsizing.

\subsection{High-level Design}
There are two types of scheduling models in ghOSt, per-CPU scheduling and centralized scheduling. In the per-CPU scheduling case, each CPU core maintains a task queue and there is an agent thread attached to each core. The agent thread is responsible for processing the message from the ghOSt kernel side and scheduling the tasks in the queue that belong to its own core. CFS is a typical per-CPU scheduler in ghOSt. The drawback of the per-CPU scheduler is that it may cause imbalances between the per-core queues.

In the centralized scheduling case, all the CPU cores maintain a global task queue and a global agent is responsible for making scheduling decisions for the whole enclave and processing the kernel message. Other agents stay inactive in the centralized model. When the global agent assigns a new task to a CPU core, the task on the core will be preempted. Shinjuku~\cite{kaffes2019shinjuku}, Shenango~\cite{ousterhout2019shenango}, and Caladan~\cite{fried2020caladan} are centralized scheduling models. 

\textbf{Two CPU Groups.} We design and implement a hybrid scheduler for serverless function scheduling by separating the scheduler into two groups of CPUs with ghOSt. Our scheduler architecture is depicted as Figure~\ref{scheduler architecture}. In our hybrid scheduler, we divide the whole ghOSt enclave into two groups of CPU cores. One group employs the FIFO policy and the other employs the CFS policy. Among the FIFO cores, the scheduler works in a centralized way and a global agent is responsible for processing the message from the kernel side. Among the CFS cores, each agent works in a per-CPU manner but is not responsible for processing messages. Messages are still processed by the first global agent.

\textbf{Short Task Group.} The first group of CPU cores is specifically used for rapidly processing short tasks whose running time is less than a configurable time limit, e.g., one can set it to 100ms. When the tasks are created, ghOSt messages are sent from the kernel side, and tasks are first directed to the global queue of the group. This group of CPU cores deploys the FIFO scheduler and schedules the tasks in a centralized way. The scheduler iterates all the cores in this group and checks if the runtime of tasks on these cores exceeds the time limit. If the task is short and will finish before running out of the pre-defined time limit, our scheduler will run it to completion. For long tasks that exceed the pre-defined time limit, the scheduler changes the status of that task, preempts it, and sends it to the second group of CPUs, the ones running longer tasks. The core running this task will simply draw another task from the global FIFO queue.

\textbf{Long Task Group.} The other group of cores in our enclave employs the CFS policy, aiming at handling the long tail of the longer-running tasks. In the CFS cores group, each core maintains an individual queue. The preempted tasks from the FIFO cores will be evenly distributed to the CFS cores in a Round-Robin way. The tasks in the CFS individual queue will be ordered by their virtual runtime. CFS policy always selects the task with the smallest virtual runtime to execute next, which ensures the fairness of among the tasks. When the task completes, a message will be sent from the ghOSt kernel side and the scheduler in the user space side frees the task process.

Through this design, with an appropriate time limit, the majority of the tasks can be completed without preemption. Thus, the number of preemption can be reduced significantly resulting in a cost reduction for the serverless users.

\subsection{CPU Cores Rightsizing and Dynamic Preemption Time Limits}
\label{sec:adaptation}

Since the design presented above improves cost for the users, we seek to not disrupt the cloud providers' operation. Combining CFS and FIFO in fixed CPU groups (e.g., 25 and 25 cores) and setting arbitrarily long time slices will result in server under utilization because one or the other group of CPUs might run out of tasks. We therefore propose two mechanisms to keep server utilization high. First, we dynamically set the FIFO preemption time limit. Second, we dynamically rightsize the two CPU groups such that neither of them remain under-utilized. 

We dynamically adapt the time limit according to the recent past task durations. In a data structure we keep the most recent 100 function durations. Using these data the scheduler chooses the time limit as a configurable percentile.

\begin{figure}[tp]
    \centering
    \includegraphics[width=0.92\linewidth]{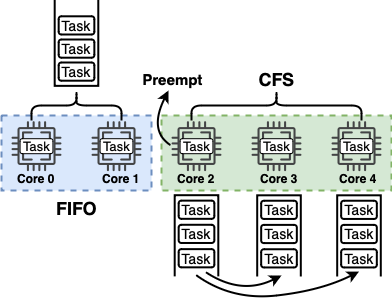}
    \caption{Core 2 migrates from CFS to FIFO. In FIFO cores group, all cores maintain a global task queue. While in CFS cores group, each core maintains queue. Tasks in the migrated core are redistributed.}
    \label{Adaptation}
\end{figure}

Figure~\ref{Adaptation} demonstrates the migration process of transferring a core from the CFS group to the FIFO group in our scheduler. Initially, Core 0 and Core 1 are dedicated to the FIFO policy and share a global queue, whereas Core 2 to Core 4 employ the CFS policy and each core maintains an individual queue. The migration process begins when we decide to transition Core 2 from the CFS group to the FIFO group. \textbf{Lock Core 2:} Before migrating, the initial step is locking the Core 2 to prevent any new task from being assigned to Core 2. \textbf{Task Preemption:} Then we check if there is a task occupying Core 2. If a task is running, it is preempted and placed into the individual queue of other Cores. \textbf{Task Migration:} Then we migrate the tasks from Core 2 queue to the remaining CFS cores (Cores 3 and 4). This redistribution aims to balance the load among the available CFS cores. \textbf{Policy Transition:} After the tasks have been successfully migrated, Core 2 is moved from the CFS group to the FIFO group. \textbf{Unlock Core 2:} Finally, Core 2 is unlocked and is available for new tasks under the FIFO policy. 

The core migration from FIFO cores to CFS cores is much easier than its counterpart. When we migrate a core from FIFO cores to CFS cores, the core will be removed from the FIFO core list and added to the CFS core list and get an individual queue. The task on that core will be preempted when we schedule a new task on that core. Since now this new core in the CFS group has an empty queue, we migrate tasks from the other cores' queues such that the queues are balanced in size.

\section{Experiment Setup}
We introduce the overview of our experiment setup and how we generate the real-world serverless function traces. Finally, we introduce the hardware and software setup.

\subsection{Overview}
\begin{figure}[tp]
    \centering
    \includegraphics[width=0.99\linewidth]{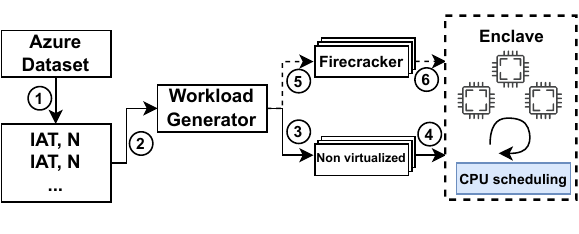}
    \caption{Our Deployment Overview: \ding{172} Extract serverless function arrival pattern from real-world Azure workload trace and generate a workload file including the inter-arrival time (IAT) and corresponding argument N of the Fibonacci function. \ding{173} Workload generator reads the items in the workload file. \ding{174} Workload generator asynchronously launches Fibonacci functions with argument N according to the corresponding IAT in the workload file. \ding{175} Get the pid of the Fibonacci process and send the pid to the ghOSt enclave. \ding{176} To measure the impact of VM virtualization, we can also launch a Firecracker microVM containing the Fibonacci workload.}
    \label{overview deployment}
    \vspace*{-0.4cm}
\end{figure}

Figure~\ref{overview deployment} illustrates the overview of our experimental deployment. Our deployment mainly consists of three parts: function trace, workload generator, and scheduler in user space. The function trace in our deployment is the Microsoft Azure trace ~\cite{shahrad2020serverless}, which contains real-world production FaaS data. This ensures that we can tune our scheduler modeling the real-wold behavior. 

To better emulate the function durations in the real-world function traces, we use Fibonacci functions with various arguments to represent functions with different durations. Fibonacci workloads are CPU and memory-bound. To benchmark scheduling algorithms, Fibonacci workloads are sufficient, other kinds of workloads would only stress other parts of the system, while the scheduler would be working in the same way. Other kinds of workloads would also likely hide possible scheduling issues due to IO pauses, adding noise to the results. Therefore in this work, we limit to the aforementioned workload. 

The workload generator extracts the serverless function arrival pattern from real-world Azure workload trace and generates a workload file containing the inter-arrival time (IAT) and corresponding argument of the Fibonacci function. The workload generator reads the items in the workload file and asynchronously launches Fibonacci functions according to the corresponding IAT in the workload file. 

Subsequently, we get the pid of the Fibonacci functions and pin the pid of the process to the ghOSt enclave. The functions are thus directed into our custom hybrid scheduler. 
We came up with two different modes of running functions: simple Unix processes and inside Firecracker~\cite{agache2020firecracker} virtual machines. This covers both scenarios of serverless running in containers and Firecracker microVMs. The former emulates the way how FaaS functions run in containers. The latter emulates the functioning of AWS Lambda. The latter also offers better isolation and security guarantees. The Firecracker analysis complements the container based usecase and shows the behavior for uVM-based platforms. However, Firecracker is a rather complex system, which spawns multiple threads for different components of the virtual machine. This makes scheduling more complex, as we will show in our experiments.

\subsection{Workload Generator}
\label{workload generator}
We discuss how to extract the arrival pattern of real-world serverless functions and how to emulate function workloads with appropriate inter-arrival time.

\textbf{Calibration.} We do a calibration for the function execution time by running the Fibonacci function binary with input $N=36..46$ for 100 repetitions, and then get the average Fibonacci function execution duration for each N. We then match these durations to function durations from the Azure traces~\cite{shahrad2020serverless}.

\textbf{Extracting Traces.} We merge the invocation per function and function duration tables from the Azure dataset~\cite{shahrad2020serverless}. We then get the average function duration and its invocation pattern per 1,440 minutes (1 day). Each row of the table has the function duration as the first item followed by 1,440 counts that refer to the invocation count in that minute. We clean the data to remove garbage (either negative or too large values). Then we group the table by the unique function duration and merge their invocation count in each minute so that in the table each row indicates the invocation count in 1,440 minutes for that unique function duration. We create buckets for which the range is the calibrated function duration in the previous step. Furthermore, we merge the rows in the table whose function duration belongs to the same bucket. Then the row of the merged table indicates the invocation time of the function with the corresponding Fibonacci N in that minute. We downscale the whole table by a factor of 100. According to this table, we can calculate the inter-arrival time for all functions.

\begin{figure}[htp]
    \centering
    \includegraphics[width=0.99\linewidth]{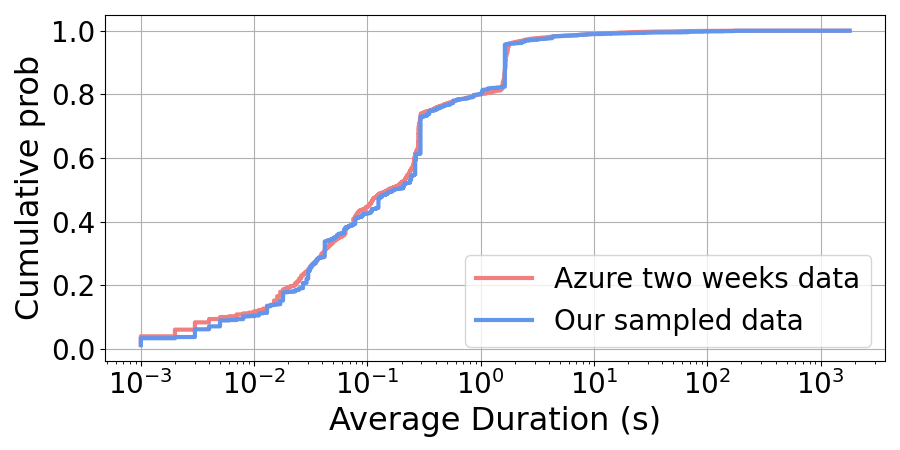}
    \caption{Two weeks of Azure traces data vs. our sampled data. The two cumulative probability function curves almost overlap, indicating our sampled data is representative.}
    \label{serverless duration compare}
    \vspace*{-0.4cm}
\end{figure}

\textbf{Workload Generation.} We pick the first two minutes of Azure trace data to generate our workload. We assume that the function arrives at regular intervals every minute. Then we can calculate the function interval time in that minute by dividing 60 by the number of function invocations in that minute. After sorting the invocations of all functions within that minute, the time difference between adjacent invocations is the inter-arrival time of each function within that minute. Figure~\ref{serverless duration compare} demonstrates the distribution similarity between our sampled data and two weeks of data from Azure trace. The two cumulative probability function curves almost overlap, indicating our sampled data is representative.

\subsection{Hardware and software setup}
\label{config}
\textbf{Hardware:} A server with two Intel Xeon CPU E5-2697 v4 @2.30GHz, each having 18 cores and 2 threads per core, with 512GB DRAM. For the workloads we use only 50 cores for the ghOSt enclave, leaving 22 cores for the OS and monitoring tasks. We believe this is standard practice in the industry.

\textbf{OS:} Ubuntu 22.04.3 LTS (GNU/Linux 5.11.0+ x86\_64) with ghOSt library installed.

\textbf{ghOSt:} cloned from ghOSt GitHub repository with the version of d4ddad4c307f2b34ee5e807e791072de29482bbe

\textbf{Firecracker:} Firecracker v1.6.0-dev

\section{Evaluation}
\label{evaluation}
Through our evaluation, we aim to answer the following:
\begin{enumerate}
    \item How does our scheduler perform with various CPU group sizes? What do we gain from our hybrid scheduler compared with FIFO and CFS policies?
    \item How do our time limit and CPU cores adaptation perform? 
    \item How does our scheduler perform with the real-world FaaS platform (Firecracker)?
\end{enumerate}

\subsection{FIFO vs. CFS core Tuning}
\begin{figure}[tp]
    \centering
    \includegraphics[width=0.99\linewidth]{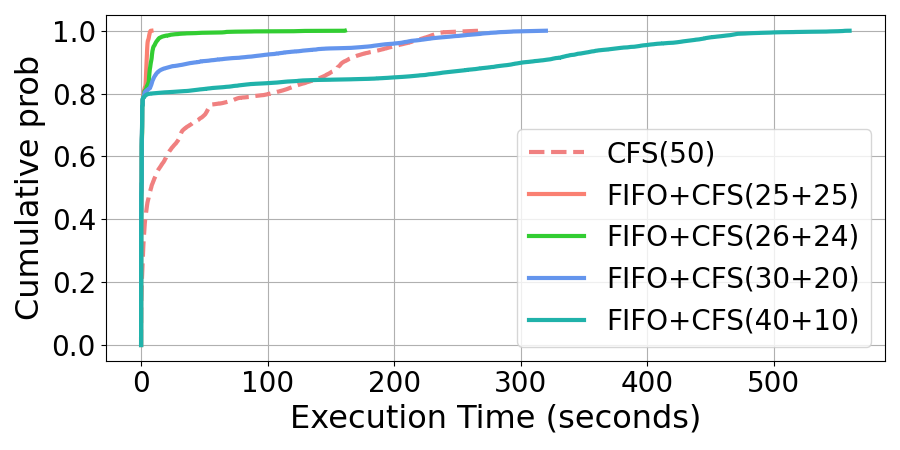}
    \caption{Execution time comparison among CFS and varies FIFO+CFS. The number in the brackets refers to the number of CPU cores designated to FIFO policy and CFS policy correspondingly.}
    \label{execution assemble}
\end{figure}

We tune the proportion of cores between the two scheduling groups. We measure the execution time of all functions processed by the enclave which contains 50 CPU cores and set the FIFO preemption time limit at 1,633 ms, which is the 90th percentile execution time from our sampled dataset. The results are shown in Figure~\ref{execution assemble}. We notice that when half of the cores run FIFO and half of the cores run CFS, the hybrid scheduler achieves the best result, where all of the functions complete execution in the first 10 seconds. Under CFS policy (the dotted line), all the functions are completed in the first 300 seconds. Note that for 40 FIFO cores 10 CFS cores, the execution CDF plot presents an obvious long tail. This is because the CFS cores are heavily utilized but the FIFO cores are under lighter load toward the tail end of the workload. 80\% of functions can be completed in a very short period but the remaining 20\% of functions are longer and have not finished in the time limit, thereby these 20\% functions are being switched among the CFS cores for a long time.

\begin{figure}[htp]
    \centering
    \includegraphics[width=0.99\linewidth]{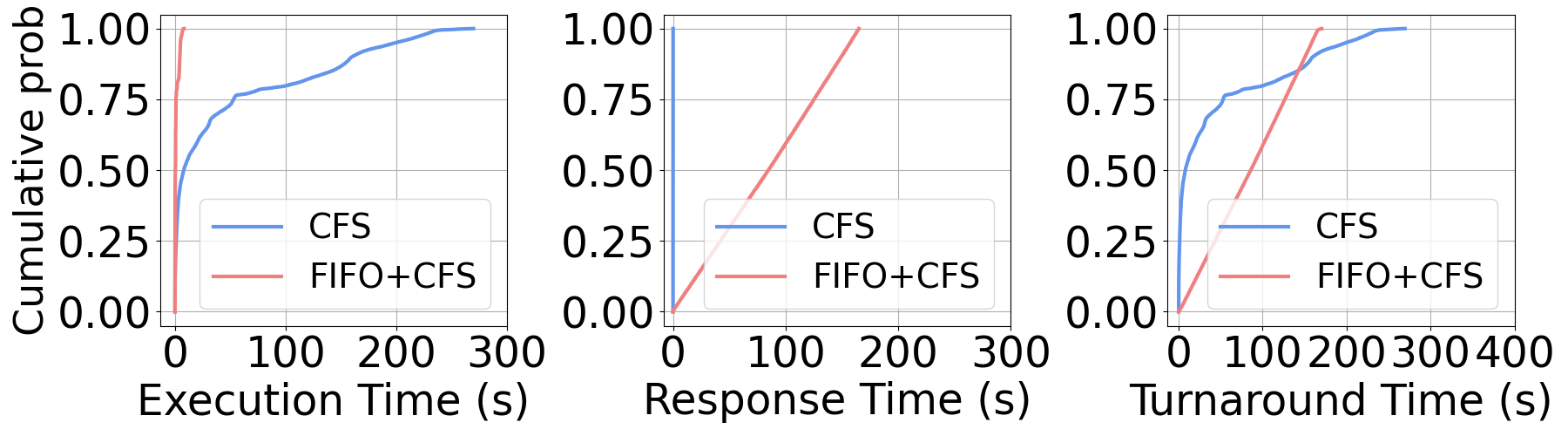}
    \caption{Metrics comparison between FIFO+CFS with CFS. Our hybrid scheduler achieves better execution time and turnaround time even though response time is worse.}
    \label{Metrics between FIFO+CFS with CFS}
\end{figure}

Figure~\ref{Metrics between FIFO+CFS with CFS} depicts the execution time, response time, and turnaround time comparison between FIFO+CFS policy with CFS policy. In the FIFO+CFS policy, we assign 25 CPU cores to the FIFO policy and remain 25 CPU cores to the CFS policy (i.e., the best configuration from Figure~\ref{execution assemble}). Figure~\ref{Metrics between FIFO+CFS with CFS} indicates that our hybrid scheduler achieves a much better execution time than the CFS policy. In terms of response time, our scheduler is worse than the CFS policy because of the nature of the CFS policy, which keeps the tasks switching to pursue fairness among tasks. Even though under our hybrid scheduler, the response time of tasks is worse, our hybrid scheduler achieves an overall better turnaround time. All tasks' turnaround time is less than 200 seconds under the hybrid scheduler. There still are $\sim$15\% tasks under CFS policy turnaround time slower than under our policy.

\begin{figure}[hp]
    \centering
    \includegraphics[width=0.99\linewidth]{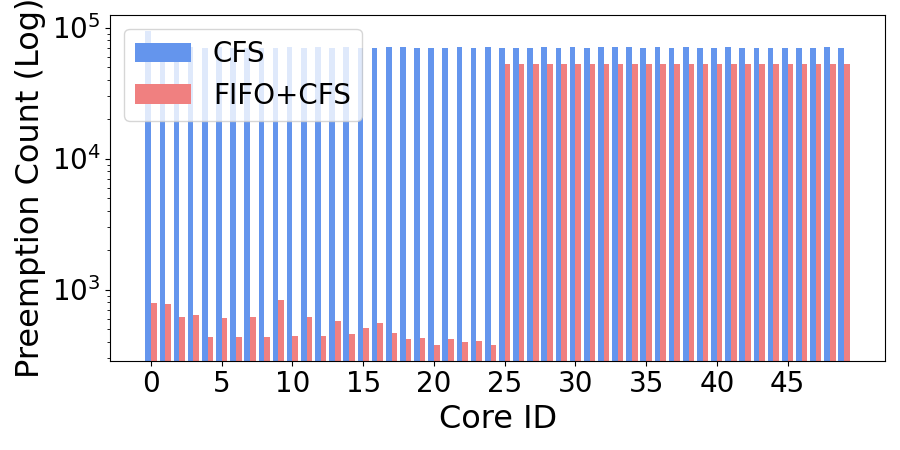}
    \caption{Preemption count per core for CFS and FIFO+CFS. First 25 CPU cores are designated FIFO policy and the remaining 25 CPU cores deploy CFS policy with 1,633 ms time limit. Note Y-axis scale is logarithmic.}
    \label{preemption count}
\end{figure}

Figure~\ref{preemption count} presents a comparison of the preemption count across all cores between our FIFO+CFS policy and CFS policy, each of them is designated 50 cores. In our FIFO+CFS policy, we allocate 25 cores for FIFO and CFS correspondingly. The plot indicates that our scheduler, which employs the FIFO policy, suffers from significantly fewer preemptions, thereby the short-lived functions can be processed seamlessly in a short period with few disruptions. The result depicts the advantages of our hybrid scheduling policy in reducing the execution time of serverless functions.

\begin{figure}[htp]
    \centering
    \includegraphics[width=0.99\linewidth]{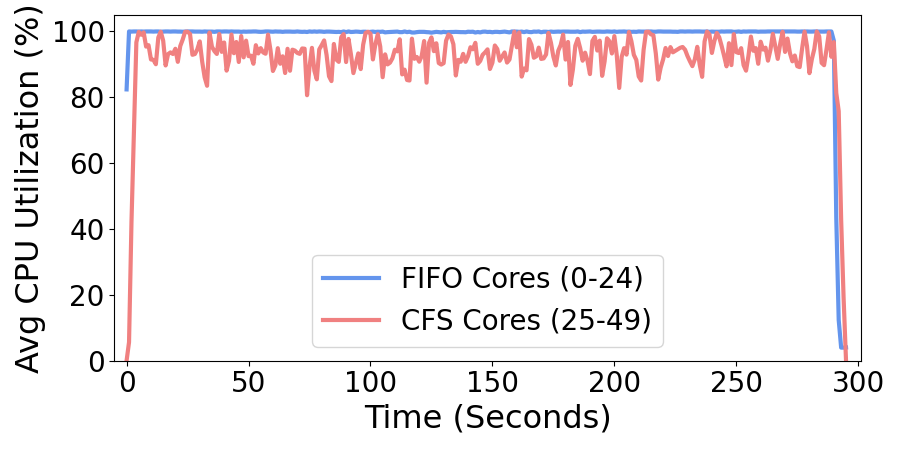}
    \caption{Average CPU utilization among cores deploying FIFO policy and cores deploying CFS policy, both policies are designated for 25 CPU cores.}
    \label{CPU utilization}
\end{figure}

Figure~\ref{CPU utilization} illustrates the average CPU utilization among FIFO cores and CFS cores, both policies are designated 25 CPU cores. The utilization of the FIFO cores as depicted is consistently high. This reflects the characteristic of the FIFO policy, where the tasks are scheduled without any interruption to their completion. CFS cores utilization fluctuates but still remains around 100\%. However, with a different preemption time limit for FIFO, utilization figures could be different.

\vspace*{0.4cm}

\noindent\fbox{%
    \parbox{\linewidth}{%
        \textbf{Conclusion 1:} Our hybrid scheduler is able to achieve good execution and turnaround times for tasks. It also reduces preemptions and can keep CPU utilization high.
    }%
}

\subsection{Preemption Time Limit Study}
\begin{figure}[htp]
    \centering
    \includegraphics[width=0.92\linewidth]{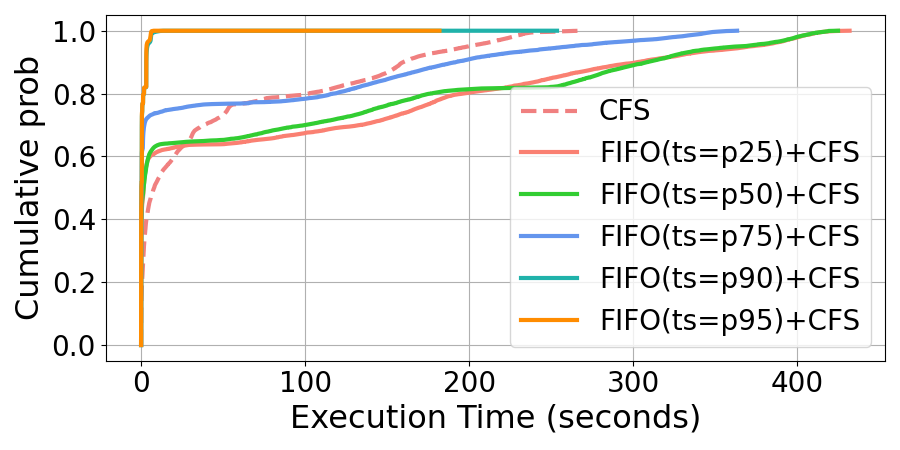}
    \caption{Execution time vs. various FIFO preemption time limits. $ts=pN$ means that the time limit allowed is equal to the Nth percentile of the previous 100 tasks.}
    \label{cdf_execution_adaptbyduration}
\end{figure}

To dynamically adapt the time limit on the FIFO cores, we set a sliding window storing the recent past 100 task execution durations. We can then choose a given percentile of these data to represent the preemption time limit. Figure~\ref{cdf_execution_adaptbyduration} shows the execution time of various time limits which correspond to 25th, 50th, 75th, 90th, and 95th percentile of the recent 100 task duration. The 95th percentile of the recent 100 task duration as depicted can achieve the best execution time. The number of FIFO and CFS cores is 25 each.

To better observe the CPU utilization of FIFO cores and CFS cores, we test our hybrid scheduler with a longer workload which consists of the tasks in first 10 minutes of the Azure trace. This is because in practical setups, in a server there will be a continuous stream of tasks coming in as depicted in Figure~\ref{spike}. 

\begin{figure}[htp]
    \centering
    \includegraphics[width=0.99\linewidth]{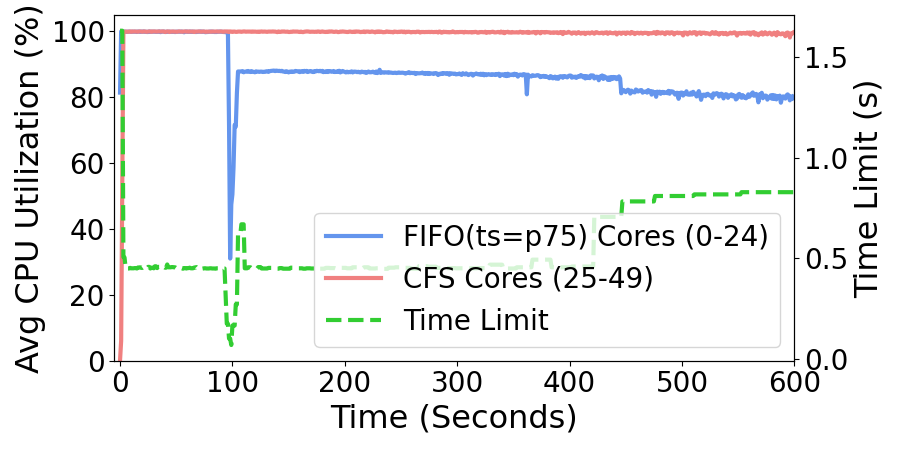}
    \caption{Average CPU utilization of the FIFO and CFS cores and time limit, when dynamically adapt the time limit according to the 75th percentile of recent 100 task durations.}
    \label{75th percentile}
\end{figure}

\begin{figure}[tp]
    \centering
    \includegraphics[width=0.99\linewidth]{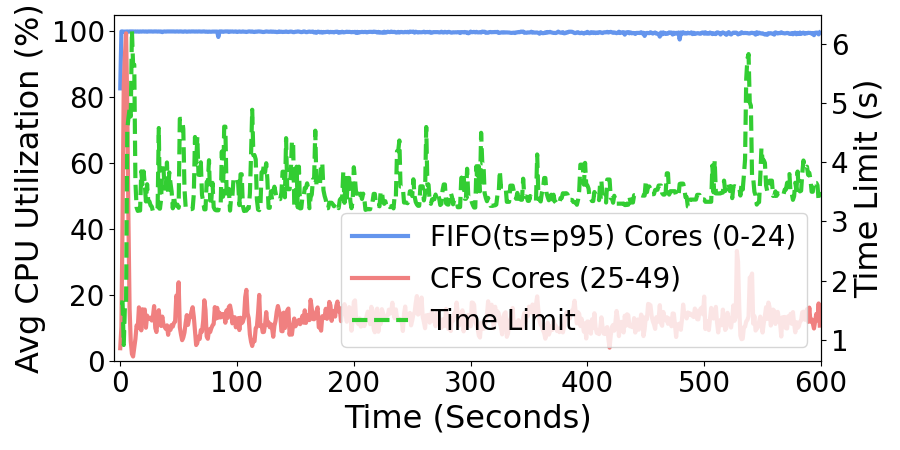}
    \caption{Average CPU utilization of the FIFO and CFS cores and time limit, when dynamically adapt the time limit according to the 95th percentile of recent 100 task durations. Note the 95th percentile of recent 100 task durations is volatile.}
    \label{95th percentile}
\end{figure}

Figure~\ref{75th percentile} and Figure~\ref{95th percentile} depict the average utilization of FIFO cores and CFS cores change over time with the changing of the time limit. In Figure~\ref{75th percentile}, the time limit adapts based on the 75th percentile of the recent 100 task durations. At the beginning, the time limit is still set as 1,633 ms and the time limit drops to 500 ms soon. After 100 seconds, the utilization of the FIFO cores stays around 90\%, indicating the time limit is too small and the tasks are preempted from the FIFO cores to the CFS cores a little too early.

In Figure~\ref{95th percentile}, the time limit is set as the 95th percentile of the recent 100 task durations. After the initial large fluctuations, the time limit hovers around 3.5 seconds. The 95th percentile of the recent 100 task durations is volatile as depicted in the plot. Because the time limit is relatively high (the 90th percentile of function durations in our workload is 1,633 ms), there are seldom tasks preempted from the FIFO cores to the CFS cores. This is not a good result for the cloud providers because it shows that there is still room to run tasks on the server, reducing the provider profit margin. A higher utilization would be better. We show how to achieve this in the next section.\\

\noindent\fbox{%
    \parbox{\linewidth}{%
        \textbf{Conclusion 2:} When adapting the time limit based on the percentile of recent task durations, the 95th percentile can achieve the best execution time of tasks among others. However, the 95th percentile of task durations is volatile and too large to properly utilize the CFS cores.
    }%
}

\subsection{CPU Policy Group Rightsizing}
As introduced in Section~\ref{sec:adaptation}, CPU cores can be migrated from one policy to another. Our goal is to achieve CPU group rightsizing such that both groups have high CPU utilization. This helps cloud providers improve profit margins by scheduling as many functions as possible and keeping machines busy, increasing overall throughput.

\begin{figure}[htp]
    \centering
    \includegraphics[width=0.99\linewidth]{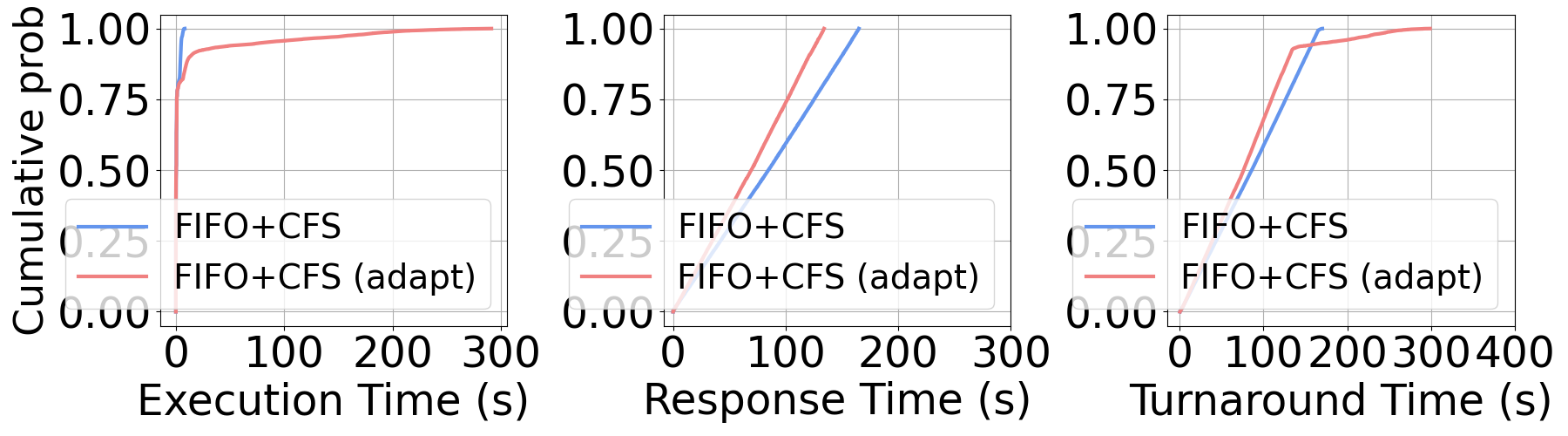}
    \caption{Metrics comparison between our hybrid scheduler with fixed cores group and with dynamically adapt cores.}
    \label{metrics for CPU adapt}
\end{figure}

We first implement a CPU utilization daemon monitoring the CPU utilization of each core through \textit{psutil}~\cite{psutil} and keep writing the CPU utilization of each core into shared memory. Our hybrid scheduler reads the CPU utilization from the shared memory and dynamically calculates the average utilization of two groups based on a time window. Then our hybrid scheduler compares the average CPU utilization of two CPU cores groups, if the difference is large enough, we then migrate one CPU core from the highly-utilized group to the under-utilized group. Through this method, our hybrid scheduler can achieve a dynamic load balance between two CPU core groups. Figure~\ref{metrics for CPU adapt} shows the metrics comparison between our hybrid scheduler with fixed cores group and with dynamically adapted cores. The scheduler with dynamically rightsized groups achieves better response time for tasks but sacrifices some execution time. 

\begin{figure}[htp]
    \centering
    \includegraphics[width=0.99\linewidth]{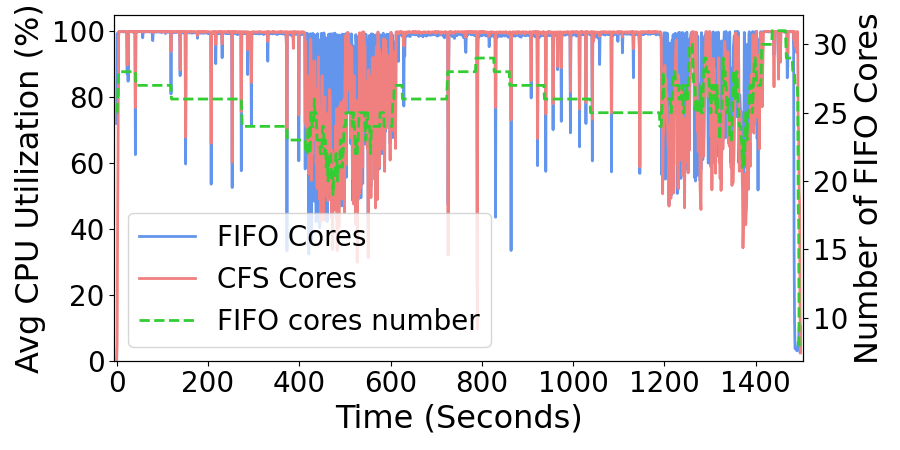}
    \caption{Average CPU utilization of the FIFO and CFS
cores and the number of the FIFO cores when adapting the CPU cores.}
    \label{Adapt the CPU group}
\end{figure}

To better observe the CPU utilization of the two changing CPU core groups, we experiment with our adaptation mechanism with a 10-minute workload to emulate cloud servers continuously running tasks. As depicted in Figure~\ref{Adapt the CPU group}, our adaption mechanism works well and the CPU utilization of two core groups stays high in general, except when migrating cores. That is expected behavior because it adds additional locking and short delays when moving tasks and cores around. This mechanism is able to help providers keep utilization high and run as many tasks as possible, improving profit margins.

\vspace*{0.4cm}

\noindent\fbox{%
    \parbox{\linewidth}{%
        \textbf{Conclusion 3:} The CPU group rightsizing of our hybrid scheduler keeps the overall CPU utilization at a high level, helping cloud providers keep a high throughput of tasks, just like with default CFS.
    }%
}

\subsection{Hybrid Scheduler Lowers FaaS Cost}
\begin{figure}[htp]
    \centering
    \includegraphics[width=0.99\linewidth]{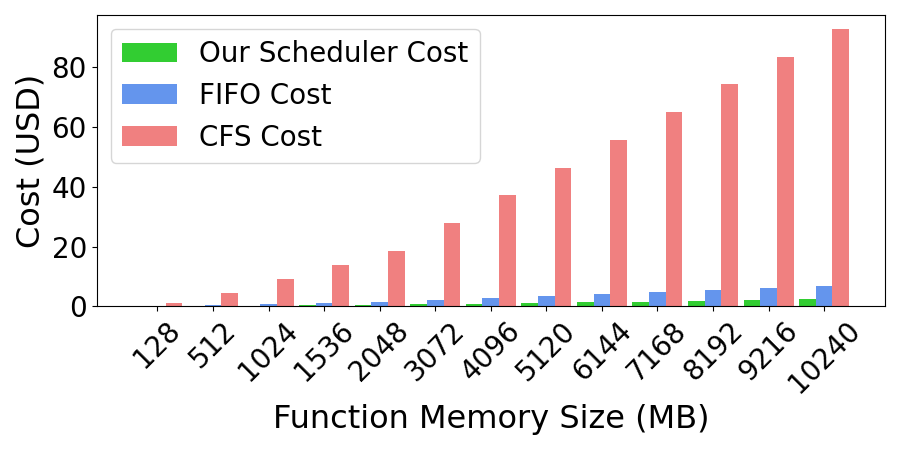}
    \caption{The cost for our hybrid scheduler, FIFO, and CFS scheduling policies calculated using AWS Lambda pricing. The workload is using the first 12,442 functions in the Microsoft Azure FaaS trace.}
    \label{cost gap our}
\end{figure}

Figure~\ref{cost gap our} shows the cost for our hybrid scheduler and CFS scheduling policies calculated using AWS Lambda pricing. We multiply function durations by the pricing for each function size and showcase what the cost difference would be if all functions would have the same size. The hybrid scheduler is able to significantly reduce user-facing costs. 

Table~\ref{tab:p99metrics} shows the p99 metrics for FIFO, CFS, and our hybrid scheduler. As the table depicts, the nature of the FIFO policy leads to a worse p99 response time. Although most tasks can achieve good execution time, the p99 execution time of FIFO in the ghOSt system suffers due to the preemption from Linux native CFS. CFS policy has the best p99 response time and can alleviate task starvation but the p99 execution time suffers from heavy task switches. Our hybrid scheduler not only reduces the p99 response time compared to FIFO, but also highly reduces the p99 execution time compared to CFS as well. In terms of cost, we calculate the overall cost according to the memory size distribution of the Azure traces among three schedulers, and using the AWS Lambda pricing~\cite{aws_pricing}. 
Our scheduler saves money for serverless consumers.

\begin{table}[ht]
    \centering
    \begin{tabular}{lccc}
        \toprule
        Metric & FIFO & CFS & Ours \\
        \midrule
        p99 Response Time (s) & 272.62 & 0.01 & 163.48 \\
        p99 Execution Time (s) & 120.16 & 232.97 & 6.69 \\
        p99 Turnaround Time (s) & 273.95 & 232.97 & 164.69 \\
        Overall Cost (USD) & 0.34 & 4.51 & 0.11 \\
        \bottomrule
    \end{tabular}
    \caption{Schedulers' overall performance and cost.}
    \label{tab:p99metrics}
\end{table}

\vspace*{0.4cm}

\noindent\fbox{%
    \parbox{\linewidth}{%
        \textbf{Conclusion 4:} Our scheduler is able to significantly
reduce user-facing costs compared with the Linux default scheduler. 
    }%
}

\subsection{Experiment with Firecracker}
All experiments above launched functions as single Linux processes, showcasing how serverless workloads work on top of containers. Several providers use virtual machines for better isolation and security instead of containers. To confirm our experiments in a realistic setting, we launch the Firecracker microVM containing the Fibonacci workload just as described before. We test the impact of the inevitable overhead of the framework (supporting various aspects of non-functional properties, from security to isolation of serverless functions) and the potential gains.

Since the overhead of launching Firecracker is more than launching a non-virtualized function it hits the limit of our server capacity much sooner. We can only launch 2,952 Firecracker microVMs, running the Azure FaaS trace. Running more microVMs does not fit in the memory of our server. However, this number is in the range of thousands of microVMs reported by AWS~\cite{agache2020firecracker} and we therefore consider this result to be realistic.

\begin{figure}[tp]
    \centering
    \includegraphics[width=0.99\linewidth]{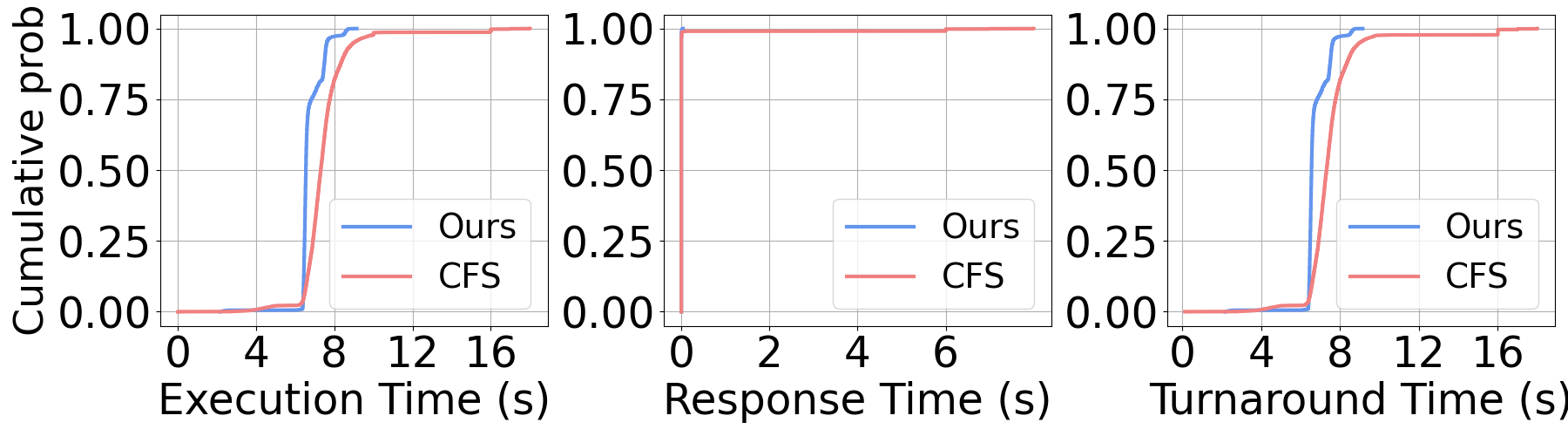}
    \caption{Launch 2,952 Firecracker microVMs. The workload is using 2,952 functions in the first 10 minutes in Microsoft Azure FaaS trace.}
    \label{firecracker}
\end{figure}

For each invocation of Firecracker microVM, there are several threads generated, each accounting for various resources of the microVM (compute, IO etc.).
We schedule all these threads under our custom ghOSt policies. We still designate 25 CPU cores for the FIFO policy and 25 CPU cores for the CFS policy. The results are depicted in Figure~\ref{firecracker}. Some microVM instances fail to launch successfully because we run out of resources. This is shown in the graph (note the horizontal line at the beginning of the execution graph). The hybrid scheduler dominates the default Linux CFS in all metrics for running our workload under Firecracker VMs.

\begin{figure}[htp]
    \centering
    \includegraphics[width=0.99\linewidth]{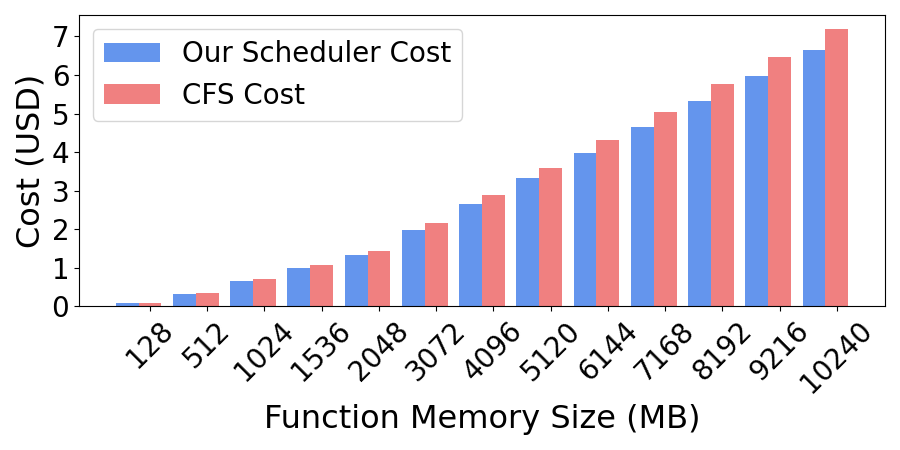}
    \caption{The cost for our hybrid scheduler and CFS scheduling policies calculated using AWS Lambda pricing. The workload is using 2,952 functions in the first 10 minutes in Microsoft Azure FaaS trace.}
    \label{cost gap firecracker}
\end{figure}

In a similar fashion to the previous cost analysis, we also compute the cost of running this workload under Firecracker and the two schedulers. Figure~\ref{cost gap firecracker} plots the results. Although the costs savings are smaller in this case, the hybrid scheduler is still able to reduce cost significantly, by a margin of about 10\%. The cost reduction is smaller in this  case because of two reasons. First, we can only schedule at most 2,952 instances before running out of memory. In the single-process case, the machine can achieve a much higher concurrency, improve cost further. Second, Firecracker is more complex, and scheduling its internal threads is not straightforward. One could envision that the internal threads of the microVM need to be scheduled according to different policies. We leave this for future work.\\

\noindent\fbox{%
    \parbox{\linewidth}{%
        \textbf{Conclusion 5:} Our hybrid scheduler is able to cost-efficiently run serverless functions on Firecracker microVMs.  
    }%
}

\section{Discussion}

\textbf{1. Making a case for better OS scheduling for serverless.} In this work we lay the groundwork for improved OS-level scheduling in serverless, showing that current defaults, such as CFS are far from achieving cost-effective schedules. Using techniques similar to ours, others could design and further experiment with (multi-level) scheduling using ghOSt. In this work, we combine FIFO and CFS, but it is outside the scope of this work to verify if this is optimal. We conjecture that in the future researchers can make use of AI-related techniques to implement better schedulers. 

\begin{figure}[tp]
    \centering
    \includegraphics[width=0.98\linewidth]{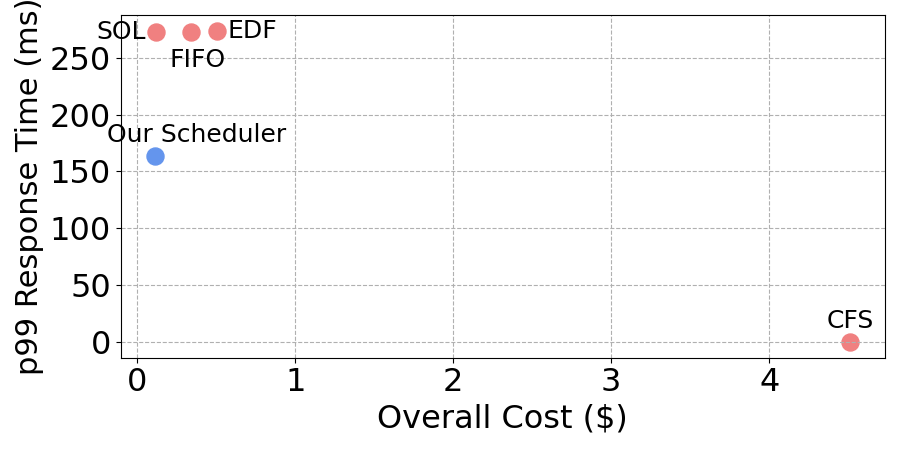}
    \caption{Cost vs p99 response time for several schedulers.}
    \label{cost vs response}
\end{figure}

As an extra exercise, we experimented with other schedulers implemented in ghOSt and plot in Figure~\ref{cost vs response} the overall cost incurred by a scheduler vs. its p99 response time. We include here our hybrid scheduler, CFS, FIFO and several others. We notice that our scheduler is close to being the best performing one in these two dimensions. We encourage the community to experiment further and design other schedulers that optimize both dimensions.

\textbf{2. The provider perspective.} In this work we have focused mostly on the user's perspective, designing a scheduler that reduces cost. On the other side of the coin, our proposed hybrid scheduler does not incur any additional overhead or nuisance for the provider due to our tailor-made preemption time limit adaptation and CPU group rightsizing. 
This ensures that cloud providers do not lose throughput compared to CFS by not keeping CPUs occupied as much as possible. 

However, it is possible that even more advantageous schemes do exist that would not only help the users but also help the providers increase the concurrency, throughput, and thus overall utilization, leading to larger margins.

\textbf{3. Containers vs. Firecracker}. We covered both containers and Firecracker microVMs in our experiments. All experiments up until the firecracker analysis also apply to containers because we spawn regular linux processes. The follow-up firecracker analysis complements this and shows the behavior for uVM-based platforms. Our results show good cost improvements in both scenarios, allowing one to conclude that our proposed hybrid scheduler is portable between virtualization stacks.

\textbf{4. Firecracker is a complex VMM.}  Firecracker, even though a lightweight virtualization solution, is a very complex system. It spawns multiple threads during runtime for performing IO, for running the actual user code~\cite{agache2020firecracker,anjali2020blending}, and, last but not least we should not forget that it runs a full Linux kernel. When scheduling it under our policies, currently all threads adhere to the same scheduling policies at any given time. This is by far not optimal, but it falls outside the scope of this work. In future work, we plan to experiment with various techniques to schedule Firecracker threads using different policies.

\textbf{5. ghOSt vs. eBPF~\cite{eBPF}}. The choice of platform here is less important as long as we achieve improved scheduling for serverless. One could have implemented the same system using eBPF. As eBPF scheduling makes its way to mainline kernels, we acknowledge that this might be the cleaner choice for the future as ghOSt needs a custom-built kernel.

\section{Related work}
We discuss related work in serverless scheduling, its impact on serverless computing, core allocation, scheduling optimizations, user-level scheduling, and cost-saving for serverless computing.

Previous work has proposed new schedulers~\cite{kaffes2019centralized, kaffes2022hermod} to suit serverless workloads better. A recent study~\cite{isstaif2023towards} modified the CFS source code showing that just modifying parameters without intervening more deeply in the way the scheduler works is insufficient. Several projects~\cite{li2023golgi, tian2022owl} overcommit serverless functions to increase server utilization and reduce the provisioning cost. The work from Zuk et al~\cite{zuk2022call} shows the scheduler impact on the response time of the FaaS system. To achieve low tail latency and high throughput when handling latency-sensitive workloads, some schedulers~\cite{ousterhout2019shenango, qin2018arachne, kaffes2019centralized, fried2020caladan, mcclure2022efficient, arutyunyan2023decentralized} use core allocation techniques. To optimize tail latency, Shinjuku~\cite{kaffes2019shinjuku} reduces the overhead of preemption, and others~\cite{prekas2017zygos,li2016work} apply work-stealing to reduce tail latency. Since it is difficult to install a kernel for a new kernel scheduler, other work~\cite{ford1996cpu,kaffes2021syrup,qin2018arachne,fu2022sfs,mvondo2022towards} proposes user-level scheduling. The work~\cite{mvondo2021tell} discusses the situation of CPU time wasted on idle units while multiple microservices are running on the same server. Several projects~\cite{eurosys23decouple,elgamal2018costless,sanchez2021experience} focus on how to reduce costs for serverless.

Closest to our work is SFS~\cite{fu2022sfs} which approximates the shortest remaining time first (SRTF) policy. SFS' drawbacks are as follows. First, SFS does not consider the cost the CFS introduces. Second, it uses schedtool~\cite{schedtool} which is limited to only a handful of policies. Our scheduler is modular enough through ghOSt such that it can use custom schedulers. Third, SFS uses OpenLambda~\cite{OpenLambda} which is a container-based serverless platform. In our experiments, we are using also Firecracker~\cite{agache2020firecracker}, which is a microVM-based platform. Scheduling microVMs is much more complex than scheduling containers because they have multiple threads working in conjunction.

\section{Conclusions}
Serverless functions are typically short-running workloads that complete in the order of seconds. To achieve economies of scale, cloud providers run thousands of these concurrently in large machines. The default scheduler in Linux, CFS, unavoidably introduces excessive preemptions thereby extending the execution time of serverless functions, and hence their cost. In this work, we proposed a hybrid scheduling method that reduces serverless functions cost. Our experiments covered both containers and Firecracker microVMs using real-world workloads show that our hybrid scheduler is capable of reducing cost by a significant margin. To the best of our knowledge, our work is the first-of-a-kind cost analysis and raising awareness for the fact that scheduling in serverless is by far not a solved problem. We present here an initial solution for reducing cost in serverless via OS-level scheduling but we acknowledge that in the future better solutions may appear. Our hybrid scheduler is open-sourced and is available at: \textcolor{blue}{\url{https://github.com/ZhaoNeil/hybrid-scheduler.git}}

\bibliographystyle{IEEEtran}
\bibliography{references.bib}

\end{document}